 \documentclass[12pt]{article}
 \usepackage{amssymb,gracos07}

 \newcommand{\beq}[1]{\begin{equation}\label{#1}}
 \newcommand{\eeq}{\end{equation}}

 \newcommand{\bear}[1]{\begin{eqnarray}\label{#1}}
 \newcommand{\ear}{\end{eqnarray}}

 \newcommand{\R}{ {\mathbb R} }
 \newcommand{\p}{\partial}

\begin{document}

\Nazvanie{5-dimensional solution  with acceleration and small
          variation of G} 

\begin{center}
\Avtor{S.B. Fadeev\Email{rikrus2005@yandex.ru}$^a$}, 
\Avtor{V.D. Ivashchuk\Email{rusgs@phys.msu.ru}$^{a,b}$}
\end{center}
\Rabota{$^a$Centre for Gravitation and Fundamental Metrology,
         VNIIMS, 46 Ozyornaya St., Moscow 119361, Russia,\\ 
   $^b$Institute of Gravitation and Cosmology,
     Peoples' Friendship University of Russia,
    6 Miklukho-Maklaya St., Moscow 117198, Russia}

\Abstract{A 5-dimensional cosmological solution in the model with two
 2-forms and  two  ``phantom'' scalar fields is
 considered. The model contains two dilatonic coupling vectors
 obeying certain restrictions.
  It is shown that there exists a time interval
 where accelerating expansion of ``our'' 3-dimensional
 space is compatible with a small  value of effective
 gravitational "constant" variation.}

 Here we study the problem of G-dot in a simple $5$-dimensional
  cosmological model with two forms of rank 2  and
  two  ``phantom'' scalar fields. (For variation of G in multidimensional
  models see \cite{IM1}-\cite{IKMN} and refs. therein).

 The main problem here is to find an interval of synchronous time
 variable $\tau$ where the scale factor of our 3-dimensional
 space exhibits
 an accelerated expansion according to observational data
 \cite{Riess,Perl} while the relative variation of the effective
 4-dimensional gravitational  constant is small enough
 in comparison   with the Hubble  parameter,
 see \cite{Hel,Dic,BZhuk} and refs. therein.

 Recently it was shown in \cite{DIKM} that in the model with two non-zero
 curvatures there  exists a time interval of $\tau$ where an accelerating 
 expansion of ``our'' 3-dimensional space is compatible with   small 
 enough value of G-dot. In this paper we suggest an analogous mechanism 
 for the model with two 2-forms and two scalar fields (e.g. phantom
 ones) when proper restrictions on dilatonic coupling vectors
 are obeyed.

 We deal here with $5$-dimensional cosmological solution in the
 model with two 2-forms  and  two scalar "phantom" fields.
 The model is  governed by the action
 \beq{1.1}
    S=\int d^5 z  \sqrt{|g|} \{
  R[g]  + \delta_{\alpha\beta} g^{MN}\p_M\varphi^\alpha
  \p_N\varphi^\beta
    - \sum_{a = 1,2} \frac{1}{2}
  \exp[2 \vec{\lambda}_a \vec{\varphi}] (F^a)^2 \}.
  \eeq
 Here  $g=g_{MN}(z)dz^M\otimes dz^N$ is
 $5$-dimensional metric
 of pseudo-Euclidean signature \\ $(-,+,+,+,+)$,
 $F^a =  dA^a$ is a 2-form,
  $\vec{\varphi} =(\varphi^1,\varphi^2) \in \R^2$ is a vector of
  two "phantom" scalar fields,
 $\vec{\lambda}_a \in \R^2$ is  dilatonic coupling vector corresponding to
  the form $F^a$, $a = 1,2$.
 In (\ref{1.1}) $|g| =   |\det (g_{MN})|$ and $\alpha, \beta = 1,2$.

   Here we consider a cosmological solution for
   the following choice of vector couplings
    \bear{1.2}
       \vec{\lambda}_1^2 =  \vec{\lambda}_2^2 = \lambda^2 >
       \frac{2}{3},
       \\ \label{1.3}
       \vec{\lambda}_1 \vec{\lambda}_2  = 1 - \frac{1}{2} \lambda^2.
      \ear

   The solution reads
   \bear{1.11}
    ds^2 = g_{MN}(z)dz^M dz^N  =
   \\ \nonumber
       X^{2A}
       \biggl\{ - dt^2 +  (dx^1)^2  + (dx^2)^2 + (dx^3)^2
         + X^{-8 h}  t^2  (dy)^2  \biggr\}, \quad
      \\  \label{1.12}
     \varphi^\alpha= - 2 h (\lambda_{1 \alpha}
    +  \lambda_{2 \alpha}) \ln X,
    \\  \label{1.13a}
    F^1= F^2 = q X^{-2} t dt  \wedge dy
   \ear

  Here $q$ is constant,
       \beq{1.5}
     X = 1 + P t^2,   \qquad    P   = \frac{1}{8h} q^2
   \eeq
    $t$ is a time variable and 
     \beq{1.14b}
       h = \frac{3}{2 - 3 \lambda^2}, \qquad A =  \frac{4 h}{3}.
      \eeq

 This solution may be verified just by a substitution into equations 
 of motion.

 In what follows we use a "synchronous"  time  variable $\tau = \tau(t)$
   \beq{2.1}
   \tau = \int_{0}^{t} d \bar{t} [X(\bar{t})]^{A}
   \eeq

 Since $\lambda^2 > \frac{2}{3}$ we get
  $P < 0$, $h < 0$ and $A < 0$.
  Let us consider two intervals  of the parameter $A$:
  \beq{2.2i}
   (a) \ A  < -1, \ {\rm or} \ 2/3 < \lambda^2 < 2
   \eeq
   and
  \beq{2.2ii}
   (b)  \ -1 < A  < 0, \  {\rm or} \  \lambda^2 > 2.
   \eeq

 For the first case $(a)$ the function $\tau = \tau (t)$
 is monotonically increasing from $0$ to $+ \infty$,
 for $t \in (0, t_1)$, where $t_1 = |P|^{-1/2}$,
 while for the second case $(b)$ it is monotonically increasing from
  $0$ to finite value $\tau_1 = \tau(t_1)$.

 The scale  factor of 3-dimensional space is
   \beq{2.3}
   a = X^A.
   \eeq

  For the first branch $(a)$ we   get an asymptotical relation
 \beq{2.4i}
   a \sim {\rm const} \  \tau^{\nu},
   \eeq
   for $\tau \to +\infty $,
 where
  \beq{2.4n}
   \nu = \frac{A}{A+1} = \frac{4}{6 -3 \lambda^2}
   \eeq
 and, due to  (\ref{2.2i}), $\nu > 1$. For the second branch
  $(b)$ we obtain
    \beq{2.4ii}
   a \sim {\rm const} \  (\tau_1 - \tau)^{\nu},
   \eeq
   for $\tau \to \tau_1 - 0$,
   where $\nu  < 0$ due to (\ref{2.2ii}).

 Thus, we are led to  an asymptotical accelerated expansion of
 3-dimensional factor space in both cases a) and  b) and
   $a \to + \infty$.

  This accelerated expansion takes place
 for all $\tau > 0$, i.e.
 \beq{2.6}
     \dot{a} > 0, \qquad   \ddot{a} > 0.
 \eeq
 Here and in what follows  we denote $\dot{f} = df/d \tau$.

 Let us prove relations (\ref{2.6}).
 Using the relation $d\tau/d t = X^A$ (see (\ref{2.1}))
 we get
 \beq{2.6a}
     \dot{a} = \frac{d t}{d \tau} \frac{d a}{d t} =
      \frac{2|A||P|t}{X} > 0,
 \eeq
 and
 \beq{2.6b}
     \ddot{a} = \frac{d t}{d \tau} \frac{d}{d t}
     \frac{da}{d \tau} =
      \frac{2|A||P|}{X^{2 + A}} (1 + |P| t^2) > 0.
 \eeq

 Now, let us consider the variation of effective gravitational constant $G$.
 For   our model  the 4-dimensional gravitational "constant"
 (in Jordan frame) is
  \beq{2.7}
       G = {\rm const} \  b^{-1} = X^{2A} t^{-1},
  \eeq
  where
   \beq{2.7a }
         b = X^{-2A} t
    \eeq
 is the scale factor of the "internal" 1-dimensional
  space with the metric $dy^2$.

   The function $G({\tau})$ has a minimum at the point
  $\tau_0$ corresponding to
  \beq{2.9}
       t_{0} = \frac{|P|^{-1}}{1 +4 |A|}.
  \eeq
  At this point the variation of $G$ is zero.  This follows from
  explicit relation for dimensionless variation of $G$
  \beq{2.11}
     \delta =  \dot{G}/(GH) = 2 + \frac{1-|P| t^2}{2 A|P| t^2}.
  \eeq
  Here
   \beq{2.12}
        H = \frac{\dot{a}}{a}
   \eeq
 is the Hubble parameter.

 The function $G({\tau})$ is monotonically
 decreasing from $+ \infty$ to $G_0 = G(\tau_0)$ for $\tau \in (0, \tau_0)$
 and  monotonically increasing from $G_0$ to $+ \infty$
 for $\tau \in (\tau_0, \bar{\tau}_1)$.
 Here $\bar{\tau}_1 = +\infty $ for the case (a) and  $\bar{\tau}_1 = 
 \tau_1$ for the case (b).

 The function $b(\tau)$ is monotonically
 increasing from zero to $b(\tau_0)$ for $\tau \in (0, \tau_0)$
 and  monotonically decreasing from $b(\tau_0)$ to zero
  for $\tau \in (\tau_0, \bar{\tau}_1)$.

 We should treat only solutions with accelerated expansion of our space
 and small enough variations of the gravitational constant obeying the
 present experimental constraint \cite{Hel,Dic}
  \beq{2.10}
        |\delta| < 0.1.
 \eeq
     Here like in the case of the model with two curvatures
    \cite{DIKM} the $\tau$   is restricted by the interval
    containing $\tau_0$.
  It follows from (\ref{2.11}) that in the asymptotical regions
  (\ref{2.4i}) and (\ref{2.4ii}) $\delta \to 2$ that is
  not acceptable by experimental bounds (\ref{2.10}). This restriction
  is satisfied for the interval containing the point $\tau_0$
  where $\delta = 0$.

     For small  $\tau - \tau_0$ we get the following approximate relation

  \beq{2.13}
       \delta  \approx  (7 + \frac{3}{2} \lambda^2) H_0(\tau - \tau_0),
  \eeq
  where $H_0 = H(\tau_0)$.  This relation gives approximate bounds
  on values of time variable $\tau$ allowed by the restriction on
  G-dot.


 {\bf Conclusions.}
 Thus, here we have considered $5D$ cosmological  solution  with two
 Abelian gauge fields  and  two phantom scalar fields.
 The solution contains  two factor spaces
 corresponding to  "our"  
 3-dimensional flat space   and to  1-dimensional "internal" space. This 
 solution takes place for special choice of dilatonic coupling vectors.
 We have found that there exists a time interval
 of $\tau$ where an accelerating expansion of ``our'' 3-dimensional
 space is compatible with   small enough value of $\dot{G}/G$
 obeying the experimental bounds. 

 {\bf Acknowlegement.} 
 This work was supported in part by the Russian Foundation for
 Basic Research grant Nr. 05-02-17478.

\end{document}